\documentclass[a4paper,twoside,twocolumn,english,superscriptaddress]{revtex4-2}
\usepackage{amsmath}
\usepackage{libertine}
\usepackage[libertine]{newtxmath}
\usepackage[LGR,T1]{fontenc}
\usepackage[latin9]{inputenc}
\usepackage{color}
\usepackage{prettyref}
\usepackage{graphicx}

\makeatletter

\pdfpageheight\paperheight
\pdfpagewidth\paperwidth

\DeclareRobustCommand{\greektext}{%
  \fontencoding{LGR}\selectfont\def\encodingdefault{LGR}}
\DeclareRobustCommand{\textgreek}[1]{\leavevmode{\greektext #1}}
\ProvideTextCommand{\~}{LGR}[1]{\char126#1}

\global\newrefformat{eq}{Eq.\,(\ref{#1})}
\global\newrefformat{fig}{Fig.\,\ref{#1}}
\usepackage[colorlinks, linkcolor=blue, anchorcolor=red, citecolor=blue]{hyperref}
\makeatother

\usepackage{babel}
\begin{document}
\global\long\def\bq{\boldsymbol{q}}%
\global\long\def\bp{\boldsymbol{p}}%
\global\long\def\bv{\boldsymbol{v}}%
\global\long\def\bB{\boldsymbol{B}}%
\global\long\def\bE{\boldsymbol{E}}%
\global\long\def\bA{\boldsymbol{A}}%
\global\long\def\d{\mathrm{\mathrm{d}}}%
\global\long\def\be{\boldsymbol{e}}%
\global\long\def\bk{\boldsymbol{k}}%
\global\long\def\bx{\boldsymbol{x}}%
\global\long\def\bxi{\boldsymbol{\xi}}%
\global\long\def\bX{\boldsymbol{X}}%
\global\long\def\ee{\mathrm{e}}%
\global\long\def\bnabla{\boldsymbol{\nabla}}%
\global\long\def\ii{\mathrm{i}}%

\title{\textcolor{magenta}{Harris Dispersion Relation and Bernstein Modes
in Dense Magnetized Quantum Plasmas}}
\author{Tian-Xing Hu}
\affiliation{Key Laboratory for Laser Plasmas and School of Physics and Astronomy,
and Collaborative Innovation Center of IFSA, Shanghai Jiao Tong University,
Shanghai, 200240, China}
\author{Dong Wu}
\email{dwu.phys@sjtu.edu.cn}

\affiliation{Key Laboratory for Laser Plasmas and School of Physics and Astronomy,
and Collaborative Innovation Center of IFSA, Shanghai Jiao Tong University,
Shanghai, 200240, China}
\author{Jie Zhang}
\email{jzhang1@sjtu.edu.cn}

\affiliation{Key Laboratory for Laser Plasmas and School of Physics and Astronomy,
and Collaborative Innovation Center of IFSA, Shanghai Jiao Tong University,
Shanghai, 200240, China}
\affiliation{Institute of Physics, Chinese Academy of Sciences, Beijing 100190,
China}
\begin{abstract}
The Bernstein wave is a well-known electrostatic eigen-mode in magnetized
plasmas, and it is of broad connection to multiple disciplines, such
as controlled nuclear fusions and astrophysics. In this work, we extend
the Bernstein mode from classical to quantum plasmas by means of the
quantum kinetic theory in a self-consistent manner, and especially
the quantum version of the Harris dispersion relation is derived.
The studied quantum effects appear in the form of pseudo-differential
operators (\textgreek{Y}DO) in the formula, which are exactly solved
using numerical methods. Furthermore, by utilizing the magnetized
equilibrium Wigner function, Landau quantization and finite temperature
effects are rigorously contained. It is found that behaviours of the
quantum Bernstein wave departure significantly from its classical
counterpart, especially when $\hbar\omega_{\mathrm{c}}$ is of the
same order of the Fermi energy. 
\end{abstract}
\maketitle

\section{\label{sec:intro}Introduction}

Bernstein modes are electrostatic eigen-modes of a magnetized plasma,
first discovered by I.B. Bernstein in 1958 \citep{PhysRev.109.10}.
Some authors \citep{Kallin1984PhysRevB.30.5655,PhysRevB.61.7517,PhysRevB.51.17744}
refer to a Bernstein mode as a ``magnetoplasmon'', since the electrostatic
eigen-mode of a unmagnetized plasma, i.e., the Langmuir wave, is often
referred to as a ``plasmon''. Bernstein modes has drawn wide interest
from multiple academic discipline. For example, electron or ion Bernstein
modes serve as an effective heating mechanism in magnetic confined
fusion researches \citep{Zhang_2012,POP2000,2006PhysRevLett.96.185003,PhysRevResearch.2.043272}.
In solid-state physics, magnetoplasmons in a 2D electron gas are a
very important topic related to semiconductor devices and quantum
Hall effect \citep{PhysRevB.30.5655,PhysRevB.83.205406,jin2016topological}.
Further more, Bernstein modes are also of interest in the fields of
space plasmas \citep{Yoon1999,Lee2018,RN51,PhysRevResearch.2.043253}
and strong coupling plasmas \citep{Bonitz2010PRL,Ott2012oscillation,Hartmann2013magnetoplasmons}.

The main objective of this paper is to extend the linear analysis
of Bernstein modes from classical plasmas to quantum plasmas. In this
context, quantum plasma refers to a plasma where the electron density
is sufficiently high that the thermal de Broglie wavelength of electrons
becomes comparable to the average distance between them, or where
the energy of a plasmon is comparable to the thermal/Fermi energy.
Quantum plasmas are ubiquitous in the universe, for example, white
dwarfs and other old or dead stars contain a significant amount of
quantum plasmas. Some models used to explain certain high-energy astrophysical
phenomena (such as X- or $\gamma$-ray bursts) also involve quantum
plasmas, and magnetic field often plays an important role in these
models. On earth, some high-energy-density experiments can produce
quantum plasma \citep{PhysRevLett.104.235003,zhang2020double,hayes2020plasma,PhysRevResearch.6.013323,10.1063/5.0211407},
and electrons in metals and electronic devices can be treated as one-component
quantum plasmas to some extent.

The linear properties of classical Bernstein modes and instabilities
have been thoroughly studied long time ago by many authors. For example,
Tataronis and Crawford \citep{tataronis1970cyclotron1} calculated
the Bernstein instabilities triggered by ring distribution and spherical
shell distribution, and they also studied oblique modes \citep{tataronis1970cyclotron2}.
However, there have been few studies on the Bernstein mode in quantum
plasmas. To the best of our knowledge, Eliasson and Shukla \citep{eliasson_numerical_2008}
were the first to study the quantum Bernstein modes in a zero-temperature
degenerate plasma, but the finite temperature and quantum diffraction
effects are neglected in their work. Iqbal etc.\,\citep{iqbal_bernstein_2017}
introduced the quantum diffraction effect by adding a Bohm potential
term into the Vlasov equation to study Bernstein modes. This methodology,
although employed by some other authors as well \citep{Tsintsadze_2009},
has not yet been verified \citep{Zheng2009comment,haas_fluid_2010}.
And, they obtained a Friedel-like oscillatory dispersion relation,
attributed to the cylindrical distribution function they employed.
The realistic distribution function of a magnetized quantum plasma
should include the effect of Landau quantization \citep{Zamanian2010},
which is also not addressed in previous works on Bernstein modes. 

The most self-consistent statistical model used to describe quantum
plasmas is the quantum kinetic theory, which reduces to the Wigner
equation in the collisionless limit. The linearization of the Wigner
equation corresponding to the so-called random phase approxiamtion
(RPA) scheme in many-body quantum physics. Relevant works using the
same method are reviewed in, e.g., Refs. \citep{Brodin2022,Manfredi2019}.

In this paper, we study the Bernstein modes by means of the linearized
Wigner-Poisson system of equations. In \prettyref{sec:Quantum-Harris-dispersion},
a general quantum Harris dispersion relation involving a pseudo-differential
operator ($\Psi$DO) in velocity space is derived. This $\Psi$DO
reflects the non-locality of quantum mechanics, which reduces to the
partial differential operator in the classical limit. In \prettyref{sec:Equilibrium-Wigner-distribution},
we introduce an exact Wigner distribution function that incorporates
Landau quantization and explain how it reduces to the non-magnetic
distribution function (Maxwellian/Fermi-Dirac) in the case of weak
magnetic fields. In \prettyref{sec:Electrostatic-modes-in}, by means
of the quantum Harris dispersion in conjunction with the Landau
quantized Wigner equilibrium function, the exact quantum Bernstein
modes are solved. 

\section{Quantum Harris dispersion relation \label{sec:Quantum-Harris-dispersion}}

Generally, motion of electrons in a quantum plasma obey the electromagnetic
quantum kinetic equation (the Wigner equation) \citep{arnold_electromagneticwigner_1989}:
\begin{equation}
\begin{aligned} & \partial_{t}f+\left\{ \frac{\bp}{m}-\frac{e}{2mc}\vartheta^{+}\left[\bA\right]\right\} \cdot\partial_{\bx}f+\frac{\ii e}{\hbar}\vartheta^{-}\left[\phi\right]f\\
 & +\frac{\ii e}{\hbar mc}\left\{ \bp\cdot\vartheta^{-}\left[\bA\right]-\frac{e}{2c}\vartheta^{-}\left[\left|\bA\right|^{2}\right]\right\} f=0,
\end{aligned}
\label{eq:EM-QKE}
\end{equation}
where $\bp=m\bv+e\bA/m$ is the canonical momentum, and the $\Psi$DO
$\vartheta^{\pm}$ is defined as
\begin{equation}
\begin{aligned}\vartheta^{\pm}\left[\phi\left(\bx\right)\right]\equiv & \phi\left(\bx+\frac{\ii\hbar}{2}\partial_{\bp}\right)\pm\phi\left(\bx-\frac{\ii\hbar}{2}\partial_{\bp}\right)\\
= & 2\begin{Bmatrix}\cos\\
-\ii\sin
\end{Bmatrix}\left(\frac{\hbar}{2}\partial_{\bx}\cdot\partial_{\bp}\right)\phi\left(\bx\right).
\end{aligned}
\end{equation}

Apply an external magnetic field $\bB_{0}=\bnabla\times\bA_{0}$ along
$z$-axis, the last term in \prettyref{eq:EM-QKE} reduce to the classical
Lorentz force term, and the brace in front of $\partial_{\bx}$ reduces
to velocity $\bv$. In an electrostatic system, the magnetic perturbation
$\delta\bB$ is ignored. Then we have

\begin{equation}
\partial_{t}f+\bv\cdot\partial_{\bx}f+\frac{\ii e}{\hbar}\vartheta^{-}\left[\phi\right]f+\frac{e}{mc}\bv\times\bB_{0}\cdot\partial_{\bv}f=0,
\end{equation}
and it is evident that a uniform static magnetic field behaves identically
in a quantum plasma as it does in a classical plasma. After linearization
to $f$ and $\phi$, we obtain

\begin{equation}
\begin{aligned}\left[\ii\left(\omega-\bk\cdot\bv\right)+\omega_{\mathrm{c}}\partial_{\varphi}\right] & \delta f=\delta\phi\frac{e}{\hbar\omega_{\mathrm{c}}}2\sinh\left(\frac{\hbar\bk\cdot\partial_{\bv}}{2m}\right)f_{0},\end{aligned}
\label{eq:linear}
\end{equation}
in the space-time Fourier space, where $\omega_{\mathrm{c}}\equiv eB_{0}/m$
is the gyro-frequency of electrons, and $\varphi$ the gyro-angle
in the cylindrical coordinate system\textbf{ }defined by \textbf{$\bv=\left(v_{x},v_{y},v_{z}\right)\equiv\left(v_{\perp}\cos\varphi,v_{\perp}\sin\varphi,v_{\parallel}\right)$.}

Integrate over both sides of \prettyref{eq:linear}, and cancel the
perturbed electrostatic potential by means of the Poisson's equation
\begin{equation}
\delta\phi=\frac{4\pi e^{2}}{k^{2}}\int\delta f\d^{3}v,
\end{equation}
we thus obtain

\begin{equation}
\begin{aligned} & 1+\frac{4\pi e^{2}}{\hbar k^{2}\omega_{\mathrm{c}}}\iiint\d v_{\parallel}v_{\perp}\d v_{\perp}\d\varphi\ee^{\ii\left(\alpha\varphi+\ensuremath{\beta}\sin\varphi\right)}\\
 & \times\int^{\varphi}\ee^{-\ii\left(\varphi'\alpha+\ensuremath{\beta}\sin\varphi'\right)}2\sinh\left(\frac{\hbar\bk\cdot\partial_{\bv}}{2m}\right)f_{\text{0}}\d\varphi'=0,
\end{aligned}
\end{equation}
where
\begin{equation}
\alpha=\frac{k_{\parallel}v_{\parallel}-\omega}{\omega_{\mathrm{c}}},\quad\beta=\frac{k_{\perp}v_{\perp}}{\omega_{\mathrm{c}}}.
\end{equation}
Noticing that
\begin{equation}
\bk\cdot\partial_{\bv}=k_{\parallel}\partial_{v_{\parallel}}+k_{\perp}\cos\varphi\partial_{v_{\perp}}-\frac{\sin\varphi}{v_{\perp}}\partial_{\varphi}.
\end{equation}
Assume the plasma is gyrotropic, then $\partial_{\varphi}=0$. After
some straightforward calculations, we finally obtain

\begin{equation}
\varepsilon\left(\omega,\bk\right)\equiv1-\frac{\omega_{\mathrm{p}}^{2}}{k^{2}}\sum_{n,\ell\in\mathbb{Z}}\int\frac{\mathcal{B}_{n\ell}\left[F_{0}\right]}{\alpha+n}\d v_{\parallel}=0,\label{eq:qhdr}
\end{equation}
where we use the uppercase letter $F=f/n_{0}$ to denote the normalized
distribution, $\omega_{\mathrm{p}}=\sqrt{4\pi e^{2}n_{0}/m}$ is the
plasma frequency, and
\begin{equation}
\begin{aligned}\mathcal{B}_{n\ell}\left[F_{0}\right]\equiv & \frac{2m}{\hbar\omega_{\mathrm{c}}}\int\mathcal{Q}_{\ell}\left(\partial_{v_{\parallel}},\partial_{v_{\perp}}\right)F_{0}\\
 & \times J_{n+\ell}\left(\frac{k_{\perp}v_{\perp}}{\omega_{\mathrm{c}}}\right)J_{n}\left(\frac{k_{\perp}v_{\perp}}{\omega_{\mathrm{c}}}\right)2\pi v_{\perp}\d v_{\perp},
\end{aligned}
\label{eq:qho}
\end{equation}
with the \textgreek{Y}DO defined by
\begin{equation}
\begin{aligned}\mathcal{Q}_{\ell}\left(\partial_{v_{\parallel}},\partial_{v_{\perp}}\right) & =\frac{1}{2}\left[\ee^{\kappa_{\parallel}\partial_{v_{\parallel}}}-\left(-1\right)^{\ell}\ee^{-\kappa_{\parallel}\partial_{v_{\parallel}}}\right]I_{\ell}\left(\kappa_{\perp}\partial_{v_{\perp}}\right)\\
 & =\begin{cases}
\sinh\left(\kappa_{\parallel}\partial_{v_{\parallel}}\right)I_{\ell}\left(\kappa_{\perp}\partial_{v_{\perp}}\right), & \ell\text{ is even,}\\
\cosh\left(\kappa_{\parallel}\partial_{v_{\parallel}}\right)I_{\ell}\left(\kappa_{\perp}\partial_{v_{\perp}}\right), & \ell\text{ is odd},
\end{cases}
\end{aligned}
\label{eq:Q-psiDO}
\end{equation}
where $\kappa_{\perp/\parallel}\equiv\hbar k_{\perp/\parallel}/2m$,
and $I_{\ell}\left(x\right)=\ii^{-\ell}J_{\ell}\left(\ii x\right)$
is the modified Bessel function of order $\ell$. In addition, the
two exponential parallel \textgreek{Y}DO terms are displacement operators
on opposite directions, hence
\begin{equation}
\begin{aligned}\mathcal{Q}_{\ell}\left[F\right]= & \frac{1}{2}I_{\ell}\left(\kappa_{\perp}\partial_{v_{\perp}}\right)\\
 & \times\left[F\left(v_{\parallel}+\kappa_{\parallel},v_{\perp}\right)-\left(-1\right)^{\ell}F\left(v_{\parallel}-\kappa_{\parallel},v_{\perp}\right)\right].
\end{aligned}
\end{equation}

In the classical limit, i.e., $\frac{\hbar\bk}{m}\cdot\partial_{\bv}\ll1$,
using 
\begin{equation}
I_{\ell}\left(x\right)\simeq\frac{1}{\ell!}\left(\frac{x}{2}\right)^{\ell},
\end{equation}
as $x\rightarrow0$, the lowest order $\ell=0$ is

\begin{equation}
\begin{aligned}\mathcal{B}_{n0}= & \frac{2m}{\hbar}\int2\pi v_{\perp}\d v_{\perp}J_{n}^{2}\left(\beta\right)\sinh\left(\kappa_{\parallel}\partial_{v_{\parallel}}\right)I_{0}\left(\kappa_{\perp}\partial_{v_{\perp}}\right)\\
= & \int2\pi v_{\perp}\d v_{\perp}J_{n}^{2}\left(\frac{k_{\perp}v_{\perp}}{\omega_{\mathrm{c}}}\right)\partial_{v_{\parallel}}+\mathcal{O}\left(\hbar\right)
\end{aligned}
\label{eq:ep0}
\end{equation}
and the second least order $\ell=\pm1$ is
\begin{equation}
\begin{aligned}\mathcal{B}_{n1}+\mathcal{B}_{n,-1}= & \frac{2m}{\hbar}\int2\pi v_{\perp}\d v_{\perp}\cosh\left(\kappa_{\parallel}\partial_{v_{\parallel}}\right)I_{1}\left(\kappa_{\perp}\partial_{v_{\perp}}\right)\\
 & \times J_{n}\left(\beta\right)\left[J_{n+1}\left(\beta\right)+J_{n-1}\left(\beta\right)\right]\\
= & \int2\pi v_{\perp}\d v_{\perp}J_{n}^{2}\left(\frac{k_{\perp}v_{\perp}}{\omega_{\mathrm{c}}}\right)\frac{n\omega_{\mathrm{c}}}{v_{\perp}}\partial_{v_{\perp}}+\mathcal{O}\left(\hbar\right)
\end{aligned}
\label{eq:ep1}
\end{equation}
where we have used the identity $J_{n+1}\left(x\right)+J_{n-1}(x)=2nJ_{n}(x)/x$.
Combining \prettyref{eq:ep0} and \prettyref{eq:ep1}, we have
\begin{equation}
\begin{aligned}\varepsilon_{c}\left(\omega,\bk\right)\equiv & 1+\frac{\omega_{\mathrm{p}}^{2}}{k^{2}}\sum_{n\in\mathbb{Z}}\iint2\pi v_{\perp}\d v_{\perp}\d v_{\parallel}\\
\times & \frac{J_{n}^{2}\left(k_{\perp}v_{\perp}/\omega_{\mathrm{c}}\right)}{\omega-k_{\parallel}v_{\parallel}-n\omega_{\mathrm{c}}}\left(k_{\parallel}\partial_{v_{\parallel}}+\frac{n\omega_{\mathrm{c}}}{v_{\perp}}\partial_{v_{\perp}}\right)F_{0}=0,
\end{aligned}
\end{equation}
which is exactly the well-known Harris dispersion relation of classical
plasmas \citep{PhysRevLett.2.34,gurnett2005introduction}. Hence,
we may refer to \prettyref{eq:qhdr} as the quantum Harris dispersion
relation, which is one of the core result of this paper.

The indices of $\mathcal{B}_{n\ell}$ range from $-\infty$ to $\infty$,
alternatively, we can define a matrix $\bar{\mathcal{B}}_{n\ell}$
with non-negative indices:
\begin{equation}
\bar{\mathcal{B}}_{n\ell}=\begin{cases}
\mathcal{B}_{00}, & n=\ell=0,\\
\mathcal{B}_{n0}+\mathcal{B}_{-n0}, & n\neq0,\ell=0,\\
2\left(\mathcal{B}_{n,\ell}+\mathcal{B}_{n,-\ell}\right), & \text{otherwise.}
\end{cases}
\end{equation}
We refer to $\bar{\mathcal{B}}_{n\ell}$ as the quantum Bernstein
matrix (QBM), which is a matrix functional of the distribution function
$f_{0}$. Then, \prettyref{eq:qhdr} becomes
\begin{equation}
\begin{aligned}\varepsilon\left(\omega,\bk\right)= & 1+\frac{\omega_{\mathrm{p}}^{2}}{k^{2}}\sum_{n,\ell\in\mathbb{Z}^{+}}\int\d v_{\parallel}\frac{\bar{\mathcal{B}}_{n\ell}\left[F_{0}\right]}{\left(\omega-k_{\parallel}v_{\parallel}\right)^{2}-n^{2}\omega_{\mathrm{c}}^{2}}\\
 & \times\begin{cases}
\left(\omega-k_{\parallel}v_{\parallel}\right)\omega_{\mathrm{c}}, & \ell\text{ is even,}\\
n\omega_{\mathrm{c}}^{2}, & \ell\text{ is odd}.
\end{cases}
\end{aligned}
\label{eq:qhdr-1}
\end{equation}

Noting that the Bessel perpendicular \textgreek{Y}DO in \prettyref{eq:Q-psiDO}
can be formulated as (see Appendix \ref{sec:on-the} for details)
\begin{equation}
\begin{aligned}I_{\ell}\left(\kappa\partial_{v}\right)F= & \frac{\left(-\right)^{\ell}}{\pi}\int_{0}^{\pi}F\left(v-\kappa\cos\varphi\right)\cos\left(\ell\varphi\right)\d\varphi\end{aligned}
.\label{eq:Jlf0}
\end{equation}
 With the aid of this formula, we can easily solve the quantum Harris
dispersion relation \prettyref{eq:qhdr} numerically, order by order.

\section{Equilibrium Wigner distribution function in a Magnetic field \label{sec:Equilibrium-Wigner-distribution}}

\begin{figure}
\includegraphics[width=0.98\columnwidth]{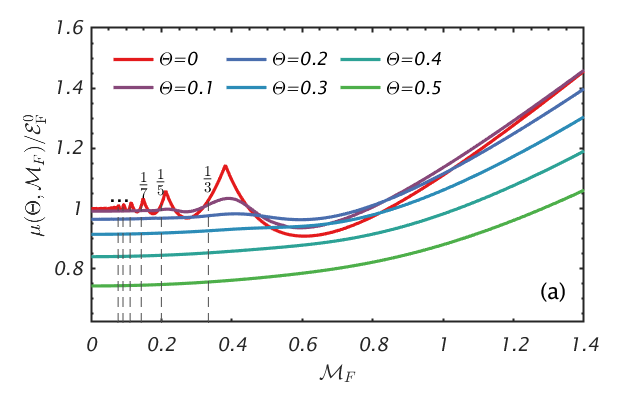}

\includegraphics[width=0.98\columnwidth]{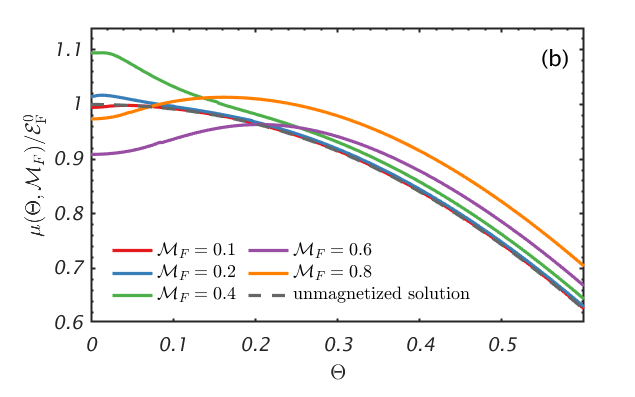}

\caption{Relations between the chemical potential $\mu$ and (a) degeneracy
$\Theta$, (b) magnetize quantum factor $\mathcal{M}_{F}$.}
\label{fig:mu}
\end{figure}
\begin{figure*}
\includegraphics[width=0.32\textwidth]{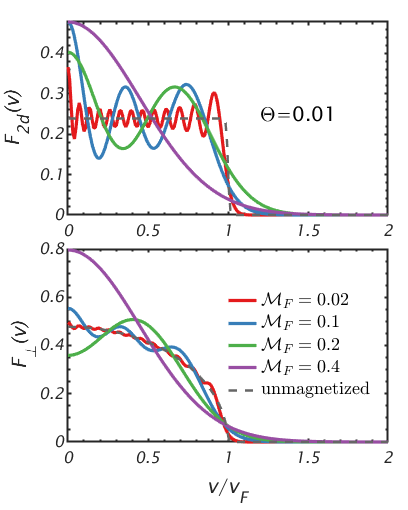}\includegraphics[width=0.32\textwidth]{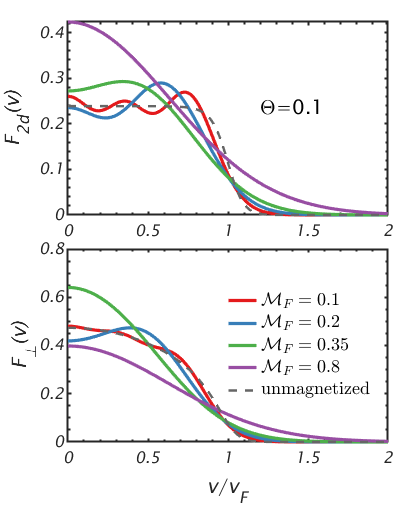}\includegraphics[width=0.32\textwidth]{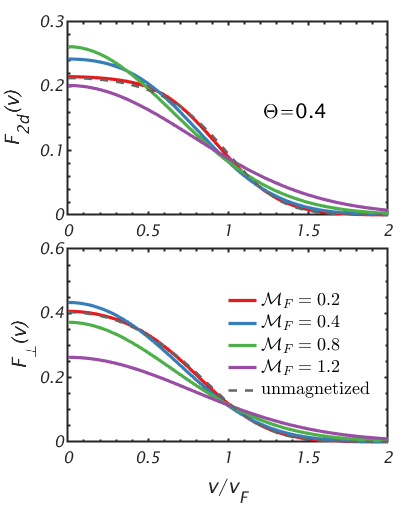}

\caption{Equilibrium Wigner function in a magnetic field.}
\label{fig:wigner}
\end{figure*}

If the magnetic field is strong enough such that $\hbar\omega_{\mathrm{c}}$
is comparable, or even much larger than the thermal or Fermi energy.
The energy of an electron is quantized into a series of Landau levels.
In such case, the equilibrium Wigner function should be (see Appendix
\ref{sec:Derivation-of-the} for details)

\begin{equation}
\begin{aligned}f_{B}\left(\bv\right) & =\mathcal{A}\sum_{n=0}^{\infty}\frac{\left(-\right)^{n}{\displaystyle 2\ee^{-\frac{mv_{\perp}^{2}}{\hbar\omega_{\mathrm{c}}}}L_{n}\left(\frac{2mv_{\perp}^{2}}{\hbar\omega_{\mathrm{c}}}\right)}}{1+{\displaystyle \ee^{\beta\left[\frac{m}{2}v_{\parallel}^{2}+\hbar\omega_{\mathrm{c}}\left(n+\frac{1}{2}\right)-\mu\right]}}},\end{aligned}
\label{eq:Wigner-B}
\end{equation}
where $L_{n}\left(x\right)$ is the Laguerre function of $n$th order.
The constant
\begin{equation}
\mathcal{A}=2\left(\frac{m}{2\pi\hbar}\right)^{3}=\frac{3n_{0}}{4\pi v_{\mathrm{F}}^{3}}
\end{equation}
appears here to convert the dimensionality of the Wigner function
into the density in the $x$-$v$ phase space, and the factor 2 stands
for fermions. Some authors prefer to alter the velocity differential
from $\d v$ to $\d p/\left(2\pi\hbar\right)$ to achieve this.

In high-temperature limit, i.e., $\beta\rightarrow0$, 
\begin{equation}
\begin{aligned}f_{B}\left(\bv\right)\simeq & \mathcal{A}\ee^{-\beta\left(\frac{m}{2}v_{\parallel}^{2}-\mu+\frac{\hbar\omega_{\mathrm{c}}}{2}\right)}\ee^{-\frac{mv_{\perp}^{2}}{\hbar\omega_{\mathrm{c}}}}\ee^{-\beta\frac{\hbar\omega_{\mathrm{c}}}{2}}\\
 & \times\sum_{n=0}^{\infty}\left(-\ee^{-\beta\hbar\omega_{\mathrm{c}}}\right)^{n}{\displaystyle L_{n}\left(\frac{2mv_{\perp}^{2}}{\hbar\omega_{\mathrm{c}}}\right)}\\
= & \frac{\mathcal{A}\ee^{\beta\mu}}{\cosh\mathcal{M}_{T}}\exp\left[-\frac{\beta m}{2}\left(v_{\parallel}^{2}+v_{\perp}^{2}\frac{\tanh\mathcal{M}_{T}}{\mathcal{M}_{T}}\right)\right],
\end{aligned}
\label{eq:wigner-classical}
\end{equation}
where we have used the identity
\begin{equation}
\sum_{n=0}^{\infty}t^{n}L_{n}\left(x\right)=\frac{\ee^{xt/\left(1-t\right)}}{1-t},\label{eq:Ln}
\end{equation}
and 
\begin{equation}
\mathcal{M}_{T}\equiv\frac{1}{2}\beta\hbar\omega_{\mathrm{c}}
\end{equation}
is the ratio between magnetic field and temperature. \prettyref{eq:wigner-classical}
is identical to the zero-order Wigner-Kirkwood expansion in a strong
magnetic field, firstly derived by Alastuey and Jancovici in Ref.\,\citep{alastuey1980magnetic}.
And noting that it reduces to Maxwellian distribution when $\mathcal{M}_{T}\rightarrow0$.
By normalizing the frequency of our system to $\omega_{\mathrm{c}}$
and velocity to the Fermi velocity $v_{\mathrm{F}}$, it is found
that the quantum effect is characterized by another dimensionless
parameter,
\begin{equation}
\mathcal{M}_{F}\equiv\frac{\hbar\omega_{\mathrm{c}}}{2\mathcal{E}_{\mathrm{F}}^{0}}=\Theta\mathcal{M}_{T},
\end{equation}
where $\mathcal{E}_{\mathrm{F}}^{0}=\hbar^{2}\left(3\pi^{2}n_{0}\right)^{2/3}/2m$
is the Fermi energy at zero magnetic field. Hence, $\mathcal{M}_{F}$
denotes the ratio between the lowest Landau level energy and the unmagnetized
Fermi energy. As $\mathcal{M}_{F}\rightarrow0$, the system also reduces
to classical.

In low-temperature limit, i.e., $\beta\rightarrow\infty$, 
\begin{equation}
f_{B}\left(\bv\right)\simeq\mathcal{A}\sum_{n=0}^{N}\left(-\right)^{n}{\displaystyle 2\ee^{-\frac{mv_{\perp}^{2}}{\hbar\omega_{\mathrm{c}}}}L_{n}\left(\frac{2mv_{\perp}^{2}}{\hbar\omega_{\mathrm{c}}}\right)},
\end{equation}
where 
\begin{equation}
N=\left\lfloor \frac{1}{\hbar\omega_{\mathrm{c}}}\left(\mathcal{E}_{\mathrm{F}}^{0}-\frac{1}{2}mv_{\parallel}^{2}\right)-\frac{1}{2}\right\rfloor 
\end{equation}
is the largest Landau quantum number that lies under the Fermi surface.
From \prettyref{eq:Ln} we have
\begin{equation}
2\ee^{-x/2}\sum_{n=0}^{\infty}\left(-1\right)^{n}L_{n}\left(x\right)=1,
\end{equation}
and since now $\hbar\omega_{\mathrm{c}}\ll\mathcal{E}_{\mathrm{F}}$,
namely, $N\gg1$, one can deduce that
\begin{equation}
2\ee^{-x/2}\sum_{n=0}^{N}\left(-1\right)^{n}L_{n}\left(x\right)\simeq\theta\left(x-\sqrt{2N}\right),
\end{equation}
then \prettyref{eq:Wigner-B} reduces to Fermi-Dirac distribution:
\begin{equation}
f_{0}\left(v_{\parallel},v_{\perp}\right)=\frac{\mathcal{A}}{1+\ee^{\beta\frac{m}{2}\left(v_{\parallel}^{2}+v_{\perp}^{2}\right)-\beta\mu}}.\label{eq:FD}
\end{equation}

In the study of Bernstein waves, the dimension parallel to the magnetic
field is usually irrelevant, which means we can integrate over $v_{\parallel}$
in advance. We define the normalized parallel and perpendicular Fermi-Dirac
function here:
\begin{equation}
\begin{aligned}F_{\parallel}\left(v_{\parallel}\right)= & n_{0}^{-1}\int_{0}^{\infty}2\pi v_{\perp}\d v_{\perp}f_{0}\left(v_{\parallel},v_{\perp}\right)\\
= & -\frac{3}{2m\beta v_{\mathrm{F}}^{3}}\mathrm{Li}_{1}\left[-\ee^{\beta\left(\mu-\frac{m}{2}v_{\perp}^{2}\right)}\right]\\
= & \frac{3}{2m\beta v_{\mathrm{F}}^{3}}\ln\left[1+\ee^{\beta\left(\mu-\frac{m}{2}v_{\parallel}^{2}\right)}\right],
\end{aligned}
\label{eq:F_para}
\end{equation}
and
\begin{equation}
\begin{aligned}F_{\perp}\left(v_{\perp}\right)= & n_{0}^{-1}\int_{-\infty}^{\infty}\d v_{\parallel}f_{0}\left(v_{\parallel},v_{\perp}\right)\\
= & -\sqrt{\frac{2\pi}{m\beta}}\mathrm{Li}_{\frac{1}{2}}\left[-\ee^{\beta\left(\mu-\frac{m}{2}v_{\perp}^{2}\right)}\right],
\end{aligned}
\label{eq:F_perp}
\end{equation}
where $\mathrm{Li}_{n}$ is the $n$-th order polylogarithm function
. In the zero-temperature limit ($\beta\rightarrow\infty$), they
become
\begin{equation}
F_{\parallel}^{0}\left(v_{\parallel}\right)=\frac{3}{4v_{\mathrm{F}}}\left(\frac{\mu}{\mathcal{E}_{\mathrm{F}}^{0}}-\frac{v_{\parallel}^{2}}{v_{\mathrm{F}}^{2}}\right),
\end{equation}
and
\begin{equation}
F_{\perp}^{0}\left(v_{\perp}\right)=\frac{3}{\pi v_{\mathrm{F}}^{2}}\sqrt{\frac{\mu}{\mathcal{E}_{\mathrm{F}}^{0}}-\frac{v_{\perp}^{2}}{v_{\mathrm{F}}^{2}}},\label{eq:0-FD-noB}
\end{equation}
since \citep{Wood92}
\begin{equation}
\mathrm{Li}_{s}\left(\pm\ee^{x}\right)\overset{x\rightarrow\infty}{\sim}-\frac{x^{s}}{\Gamma\left(s+1\right)}.
\end{equation}
Similarly, from \prettyref{eq:Wigner-B} we have
\begin{equation}
\begin{aligned}F_{B\perp}\left(v_{\perp}\right)= & n_{0}^{-1}\int_{-\infty}^{\infty}\frac{\d v_{\parallel}}{2\pi\hbar/m}f_{B}\left(v_{\parallel},v_{\perp}\right)\\
= & -\ee^{-\frac{mv_{\perp}^{2}}{\hbar\omega_{\mathrm{c}}}}\sum_{n=0}^{\infty}\left(-\right)^{n}L_{n}\left(\frac{2mv_{\perp}^{2}}{\hbar\omega_{\mathrm{c}}}\right)\\
 & \times\sqrt{\frac{\Theta}{\pi}}\mathrm{Li}_{\frac{1}{2}}\left\{ -\ee^{\beta\left[\mu-\hbar\omega_{\mathrm{c}}\left(n+\frac{1}{2}\right)\right]}\right\} ,
\end{aligned}
\label{eq:FB_perp}
\end{equation}
and in zero-temperature limit, it is

\begin{equation}
\begin{aligned}F_{B\perp}^{0}\left(v_{\perp}\right)= & n_{0}^{-1}\int_{-\infty}^{\infty}f_{B}\left(v_{\parallel},v_{\perp}\right)\d v_{\parallel}\\
= & -2\sqrt{2}\ee^{-\frac{mv_{\perp}^{2}}{\hbar\omega_{\mathrm{c}}}}\sum_{n=0}^{\infty}\left(-\right)^{n}L_{n}\left(\frac{2mv_{\perp}^{2}}{\hbar\omega_{\mathrm{c}}}\right)\\
 & \times\sqrt{\frac{\mu}{\mathcal{E}_{\mathrm{F}}^{0}}-\mathcal{M}_{F}\left(2n+1\right)}.
\end{aligned}
\label{eq:FB_perp-1}
\end{equation}

Noticing that, if the Fermi energy (or, more generally, the chemical
potential) does not vary with $\mathcal{M}_{F}$, then the lowest
Landau state has an energy higher than the Fermi energy when $\mathcal{M}_{F}>1$,
which is unphysical. Hence, we must assume $\mu=\mu\left(\Theta,\mathcal{M}_{F}\right)$
and $\mathcal{E}_{\mathrm{F}}=\mathcal{E}_{\mathrm{F}}\left(\mathcal{M}_{F}\right)\equiv\mu\left(0,\mathcal{M}_{F}\right)$.
One can determine $\mu$ by solving the equation
\begin{equation}
\frac{3}{4}\mathcal{M}_{F}\sqrt{\frac{\pi}{\Theta}}\sum_{n=0}^{\infty}\mathrm{Li}_{\frac{1}{2}}\left\{ -\ee^{\left[\frac{\mu}{\mathcal{E}_{\mathrm{F}}^{0}}-\mathcal{M}_{F}\left(2n+1\right)\right]/\Theta}\right\} =1,\label{eq:mu_Theta}
\end{equation}
given $\Theta$ and $\mathcal{M}_{F}$, or simply
\begin{equation}
3\mathcal{M}_{F}\sum_{n=0}^{N}\sqrt{\frac{\mathcal{E}_{\mathrm{F}}}{\mathcal{E}_{\mathrm{F}}^{0}}-\mathcal{M}_{F}\left(2n+1\right)}=1,
\end{equation}
when $\Theta\rightarrow0$. Furthermore, when $\mathcal{M}_{F}\rightarrow0$,
the solution of \prettyref{eq:mu_Theta} should be identical to the
unmagnetized chemical potential $\mu\left(\Theta\right)$, namely,
the solution of
\begin{equation}
\int_{0}^{\infty}\frac{4\pi v^{2}\d v}{1+\ee^{\left(v^{2}/v_{\mathrm{F}}^{2}-\mu/\mathcal{E}_{\mathrm{F}}^{0}\right)/\Theta}}=\mathcal{A}^{-1}.\label{eq:nonmagmu}
\end{equation}

Relations between $\mu$ and $\Theta,\mathcal{M}_{F}$ are plotted
in \prettyref{fig:mu}. In \prettyref{fig:mu} (a), one can see that
the $\Theta=0$ solution exhibits multiple peaks, each corresponding
to a coincidence of a Landau level and the Fermi energy, i.e., $\mathcal{E}_{\mathrm{F}}/\mathcal{E}_{\mathrm{F}}^{0}-\mathcal{M}_{F}\left(2n+1\right)$.
Since $\mathcal{E}_{\mathrm{F}}/\mathcal{E}_{\mathrm{F}}^{0}\sim1$,
the peaks are located around $1/(2n+1)$, with $n\in\mathbb{Z}^{+}$.
When $\mathcal{M}_{F}$ is greater than $0.38\gtrsim1/3$, corresponding
to the position of the last peak, there exists only one Landau level
below the Fermi surface, which is the ground state $\hbar\omega_{\mathrm{c}}/2$.
As $\mathcal{M}_{F}$ decreases, more and more Landau levels fall
below the Fermi surface. This peaky behavior of Fermi energy in a
magnetic field is well-known in solid-state physics area \citep{RN52,influence2022},
but is rarely discussed in plasma physics research. Also, the peaks
disappear when $\Theta>0.1$ , which means that this phenomenon only
occurs at extremely low temperature. In \prettyref{fig:mu} (b), the
chemical potential is presented as the function of $\Theta$, one
can see that when $\mathcal{M}_{F}$ is small, the solution does reduce
to the solution of \prettyref{eq:nonmagmu} (the dashed-line).

Equilibrium Wigner functions with varies of parameters are plotted
in \prettyref{fig:wigner}, where $F_{2d}$ is the equilibrium distribution
in a 2D system, namely, \prettyref{eq:Wigner-B} with $v_{\parallel}=0$,
and $F_{\perp}$ is defined in \prettyref{eq:FB_perp}. In the $\Theta=0.01$
panel, when $\mathcal{M}_{F}<0.1$, the Wigner function is highly
oscillatory around the unmagnetized Fermi-Dirac distribution curve,
and when $\mathcal{M}_{F}>0.1$, the magnetized curves already differ
significantly from the Fermi-Dirac curve. From the $\Theta=0.1$ and
$0.4$ panel, one can see that as the temperature increases, it takes
a higher $\mathcal{M}_{F}$ for the Wigner function to deviate from
Fermi-Dirac shape. Moreover, when temperature is high enough, the
role of $\mathcal{M}_{F}$ is to broaden the velocity distribution,
just as \prettyref{eq:wigner-classical} suggested.

\section{Electrostatic modes in magnetized quantum plasmas\label{sec:Electrostatic-modes-in}}

\subsection{Parallel propagation: Langmuir wave}

From \prettyref{eq:Q-psiDO}, we can see that if $k_{\perp}=0$, only
$\mathcal{B}_{00}$ is non-zero, since both $J_{\ell}\left(0\right)$
and $I_{\ell}\left(0\right)$ have finite values only when $\ell=0$.
\prettyref{eq:qhdr} then reduce to the RPA eigen-equation:
\begin{equation}
\begin{aligned}\varepsilon\left(\omega,k_{\parallel}\right)= & 1+\frac{m\omega_{\mathrm{p}}^{2}}{\hbar k_{\parallel}^{2}}\int\frac{F_{\parallel}^{+}-F_{\parallel}^{-}}{\omega-k_{\parallel}v_{\parallel}}\d v_{\parallel}=0,\end{aligned}
\label{eq:RPA}
\end{equation}
where $F_{\parallel}^{\pm}=F_{\parallel}\left(v_{\parallel}\pm\kappa_{\parallel}\right)$.
\prettyref{eq:RPA} corresponds to a quantum Langmuir wave \citep{hu2022kinetic}.
This is easy to understand since the electron wave along the magnetic
field line is equivalent to a Langmuir wave in an electrostatic system.
The analytic behaviors of \prettyref{eq:RPA} are thoroughly discussed
in Ref. \citep{Hu2024PRE}.

\subsection{Oblique Propagation}

Oblique modes involve both even and odd $\ell$ modes, and finite
$k_{\parallel}$ would lead to dissipative-type solutions, which makes
it more complicated than the perpendicular modes. Oblique modes are
left for future work.

\subsection{Perpendicular propagation: Bernstein modes}

\begin{figure}
\begin{centering}
\includegraphics[width=0.98\columnwidth]{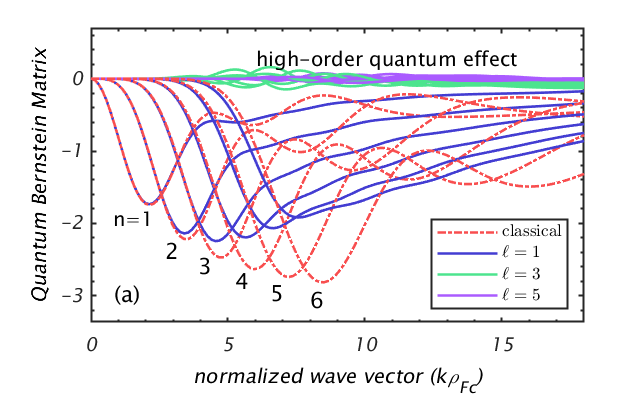}
\par\end{centering}
\begin{centering}
\includegraphics[width=0.98\columnwidth]{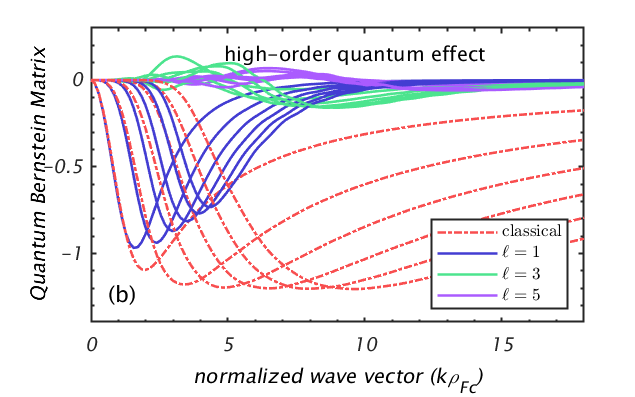}
\par\end{centering}
\caption{Zero-temperature QBM at (a) $\mathcal{M}_{F}=0.1$ and (b) $\mathcal{M}_{F}=0.8$.
The curves labeled ``classical'' are calculated for Fermi-Dirac
distribution without quantum wave effect. The subscript $n$ of some
curves has been omitted for the sake of clarity.}
\label{fig:QBM}
\end{figure}

Consider those modes propagating perpendicular to the magnetic field,
i.e., $k_{\parallel}=0$. Then the quantum Harris dispersion relation
reduces to

\begin{equation}
\varepsilon\left(\omega,k_{\perp}\right)\equiv1+\frac{\omega_{\mathrm{p}}^{2}}{k_{\perp}^{2}}\sum_{n\in\mathbb{Z}^{+}}\frac{n\omega_{\mathrm{c}}^{2}}{\omega^{2}-n^{2}\omega_{\mathrm{c}}^{2}}\sum_{\ell\in2\mathbb{Z}^{+}+1}\bar{\mathcal{B}}_{n\ell}\left[F_{\perp}\right],\label{eq:bernstein-dr}
\end{equation}
which is the eigen-equation of the quantum Bernstein modes. Now the
QBM becomes
\begin{equation}
\begin{aligned}\bar{\mathcal{B}}_{n\ell}= & \frac{4m}{\hbar\omega_{\mathrm{c}}}\int2\pi v_{\perp}\d v_{\perp}J_{n}\left(\frac{k_{\perp}v_{\perp}}{\omega_{\mathrm{c}}}\right)\\
 & \times\left[J_{n+\ell}\left(\frac{k_{\perp}v_{\perp}}{\omega_{\mathrm{c}}}\right)+J_{n-\ell}\left(\frac{k_{\perp}v_{\perp}}{\omega_{\mathrm{c}}}\right)\right]I_{\ell}\left(\kappa_{\perp}\partial_{v_{\perp}}\right).
\end{aligned}
\label{eq:QBM_b}
\end{equation}
In the classical limit, we can keep only the lowest order, and further
reduce $I_{1}\left(x\right)$ to $x/2$, then

\begin{equation}
\begin{aligned}\bar{\mathcal{B}}_{n1}= & 2n\int2\pi v_{\perp}\d v_{\perp}J_{n}^{2}\left(\frac{k_{\perp}v_{\perp}}{\omega_{\mathrm{c}}}\right)\partial_{v_{\perp}}+\mathcal{O}\left(\hbar\right)\end{aligned}
.\label{eq:QBM_b-1}
\end{equation}

Numerical results for the components of QBM are presented in \prettyref{fig:QBM},
where $\mathcal{M}_{F}=0.1$ for panel (a) and $\mathcal{M}_{F}=0.8$
for panel (b). The wave number is normalized to inverse of the radio
between Fermi velocity and gyro-frequency, $\rho_{\mathrm{F}c}=v_{\mathrm{F}}/\omega_{\mathrm{c}}$.
The red-dashed-lines, labeled ``classical'', are calculated from
\prettyref{eq:QBM_b-1} with Fermi-Dirac distribution, i.e., they
only include degenerate effects, and exclude quantum wave effects.
As is shown in both panel (a) and (b), the ``classical'' results
only coincide with the $\ell=1$ quantum modes at small $k\rho_{\mathrm{F}c}$.
The high-order quantum modes, which are the components in QBM with
$\ell>1$, have minimal effect in the $\mathcal{M}_{F}=0.1$ case,
but their impact is significant when $\mathcal{M}_{F}=0.8$. One can
see that when $k\rho_{\mathrm{F}c}\gtrsim7$, the absolute values
of $\ell=3$ modes become larger than the $\ell=1$ modes.

Substitute \prettyref{eq:QBM_b-1} into \prettyref{eq:bernstein-dr},
we have
\begin{equation}
\begin{aligned}\varepsilon\left(\omega,k_{\perp}\right)= & 1+\frac{\omega_{\mathrm{p}}^{2}}{k_{\perp}^{2}}\sum_{n\in\mathbb{Z}^{+}}\frac{2n^{2}\omega_{\mathrm{c}}^{2}}{\omega^{2}-n^{2}\omega_{\mathrm{c}}^{2}}\\
 & \times\int2\pi\d v_{\perp}J_{n}^{2}\left(\frac{k_{\perp}v_{\perp}}{\omega_{\mathrm{c}}}\right)\partial_{v_{\perp}}F_{\text{\ensuremath{\perp}}}=0.
\end{aligned}
\label{eq:bernstein-semiclassical}
\end{equation}
If the temperature is high enough such that $\Theta\gg1$, then $F_{\perp}$
tends to Maxwellian, and \prettyref{eq:bernstein-semiclassical} yields
the classical Bernstein modes \citep{PhysRev.109.10,gurnett2005introduction}:
\begin{equation}
\varepsilon\left(\omega,k_{\perp}\right)=1-\frac{2\omega_{\mathrm{p}}^{2}}{\beta_{\mathrm{c}}^{2}\omega_{\mathrm{c}}^{2}}\sum_{n\in\mathbb{Z}^{+}}\frac{\exp\left(-\beta_{\mathrm{c}}^{2}\right)}{\left(\omega/n\omega_{\mathrm{c}}\right)^{2}-1}I_{n}\left(\beta_{\mathrm{c}}^{2}\right),\label{eq:bernstein-classical}
\end{equation}
where $\beta_{\mathrm{c}}\equiv k_{\perp}v_{\mathrm{th}}/\omega_{\mathrm{c}}$. 

\begin{figure*}
\includegraphics[width=0.48\textwidth]{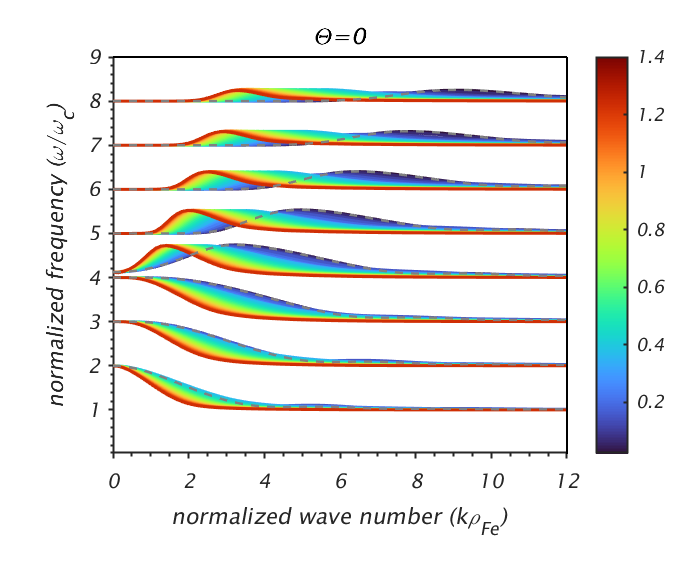}\includegraphics[width=0.48\textwidth]{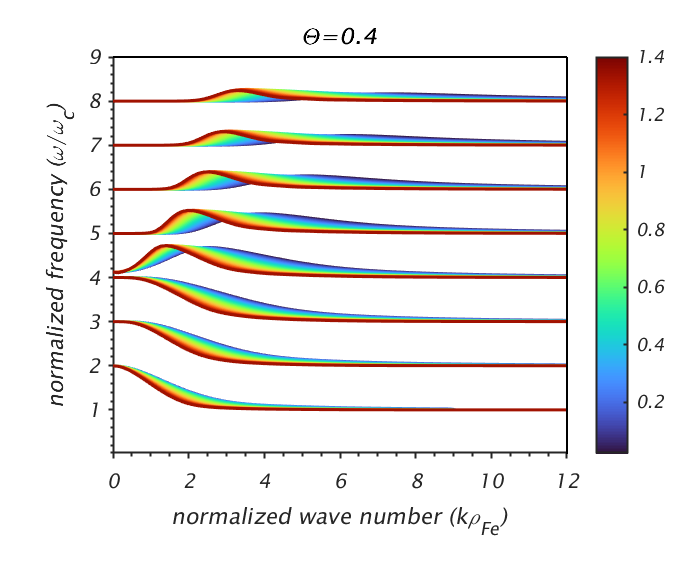}

\caption{Quantum Bernstein modes ($\omega_{\mathrm{p}}/\omega_{\mathrm{c}}=4$)
of cold ($\Theta=0$) and warm ($\Theta=0.4$) plasmas, with $\mathcal{M}_{F}=0\sim1.4$.
The dashed-lines are the solutions of \prettyref{eq:ES}, which is
obtained by Eliasson and Shukla in Ref.\,\cite{eliasson_numerical_2008}.}
\label{fig:bernstein}
\end{figure*}

Noticing that when the \textgreek{Y}DO reduces to a differential operator,
the summation over $n$ can be avoided in light of the properties
of Bessel functions \citep{tataronis1970cyclotron1,tataronis1970cyclotron2,eliasson_numerical_2008}.
It is difficult to use this scheme when a \textgreek{Y}DO is involved.
Fortunately, the high-$n$ components have little effect on the solution
and the result converges very quickly. In Ref.\,\citep{eliasson_numerical_2008},
the dispersion relations of Bernstein modes are obtained from the
eigen-equation:
\begin{equation}
1+\frac{3\omega_{\mathrm{p}}^{2}}{\omega_{\mathrm{c}}^{2}}\int_{0}^{\pi}\d\varphi\frac{\sin\varphi\sin\left(\varphi\Omega\right)}{\xi^{3}\sin\left(\pi\Omega\right)}\left(\sin\xi-\xi\cos\xi\right)=0,\label{eq:ES}
\end{equation}
where $\Omega=\omega/\omega_{\mathrm{c}}$ and $\xi=2k_{\perp}\rho_{\mathrm{F}c}\cos\left(\varphi/2\right)$.
\prettyref{eq:ES} is equivalent to \prettyref{eq:bernstein-semiclassical}
with zero-temperature Fermi-Dirac distribution \prettyref{eq:0-FD-noB}.
As mentioned in the introduction, their work neglects finite temperature
and quantum wave effects. Hence, when both $\Theta$ and $\mathcal{M}_{F}$
tend to 0, our solutions should reduce to the solution of \prettyref{eq:ES},
which are marked by the dashed-lines in \prettyref{fig:bernstein}.
Typical solutions of quantum Bernstein modes are plotted in \prettyref{fig:bernstein}.
One can see that all branches of the dispersion relation shift towards
smaller wave numbers due to quantum effects, with the high-$n$ branches
exhibiting greater shift. And, the quantum effect is quite significant
in quantum plasmas. For example, the corresponding wave number of
the maximum frequency of the $n=8$ branch shrank by almost a factor
of 3 from $\mathcal{M}_{F}=0$ to $1.4$.

\section{Discussion and Summary}

The quantum mechanical version of the Harris dispersion relation is
derived in this paper, and the non-local pseudo-differential operators
arising from quantum recoil effects are addressed self-consistently.
The effects of Landau quantization are also considered by introducing
the magnetized Wigner equilibrium function. As a example, quantum
Bernstein modes under this framework are numerically calculated. 

\begin{figure}
\includegraphics[width=0.98\columnwidth]{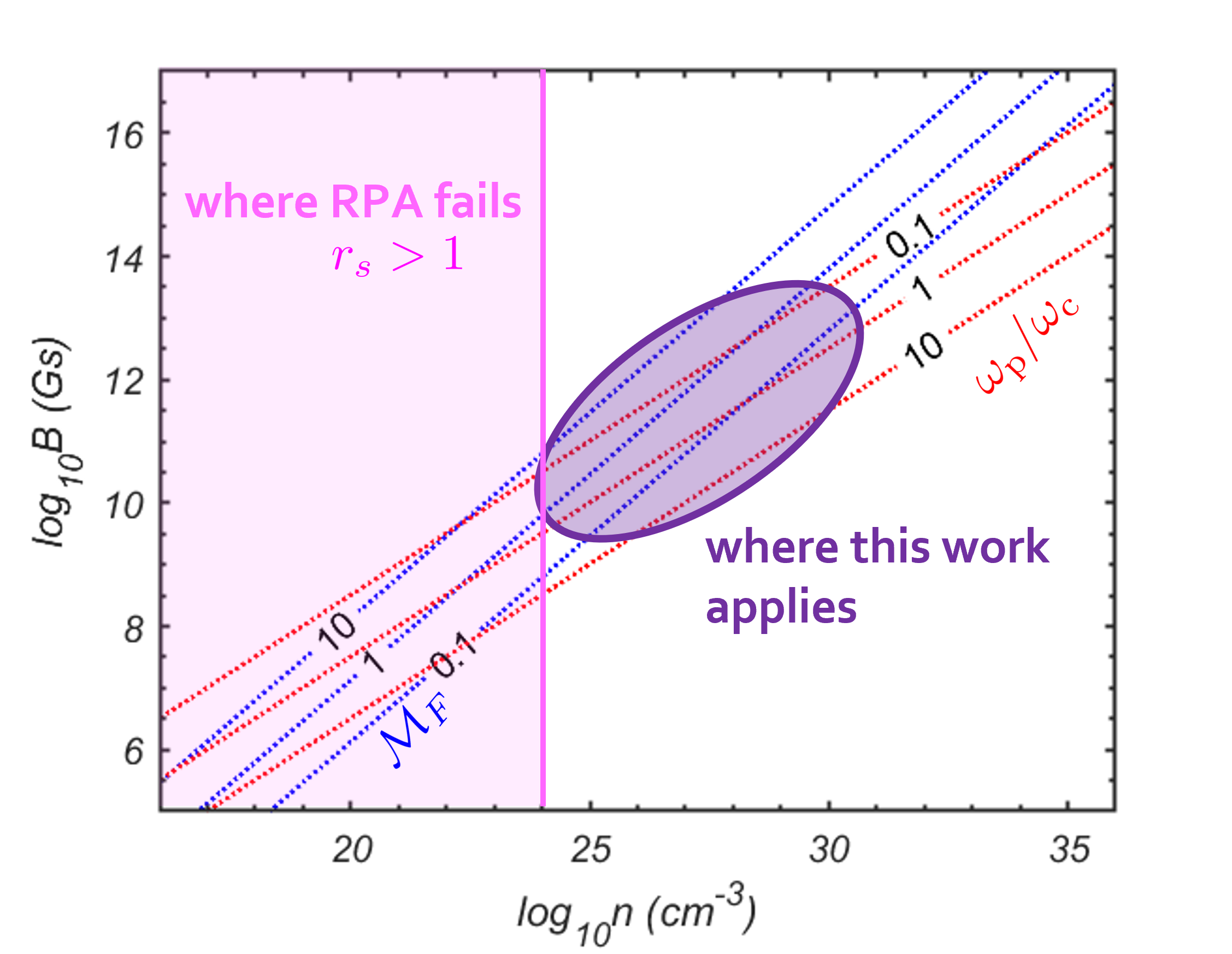}

\caption{The applicability of this paper generally covers cases where both
$\mathcal{M}_{F}$ and $\omega_{\mathrm{p}}/\omega_{\mathrm{c}}$
are of the order of 1, and RPA remains valid.}
\label{fig:parameter}
\end{figure}

All the results obtained in the main text are in normalized units,
meaning they can be applied to any density, temperature and magnetic
field strength. However, a Bernstein mode is only important when $\omega_{\mathrm{p}}$
and $\omega_{\mathrm{c}}$ are of the same order. Additionally, in
our paper, $\mathcal{M}_{F}$ also should be in order 1, as quantum
effects can be neglected if it is too small. We marked the range of
magnetic field and electron number density of this paper as a purple
ellipse in \prettyref{fig:parameter}, which is roughly where $\mathcal{M}_{F}=0.1\sim10$
and $\omega_{\mathrm{p}}/\omega_{\mathrm{c}}=0.1\sim10$ overlaps.
Note that the ellipse does not extend into the region where $n\lesssim10^{24}\mathrm{cm}^{-3}$,
which corresponds to where the Wigner-Seitz radius $r_{s}>1$, i.e.,
where the RPA breaks down. Such strong magnetic field strength ($10^{9}\sim10^{13}\mathrm{G}$)
are quite common in extreme astrophysics. For example, the magnetosphere
of a pulsar \citep{philippov_pulsar_2022}, or a high field magnetic
white dwarf \citep{Ferrario2015,Tremblay_2015}. And the electron
number density range, $10^{24}\sim10^{30}\mathrm{cm}^{-3}$, also
encompasses non-relativistic white dwarfs or accretion layers of compact
objects such as a black holes or a neutron stars.
\begin{acknowledgments}
This work was supported by the Strategic Priority Research Program
of Chinese Academy of Sciences (Grant No. XDA250050500), the National
Natural Science Foundation of China (Grant No. 12075204), and Shanghai
Municipal Science and Technology Key Project (No. 22JC1401500). Dong
Wu thanks the sponsorship from Yangyang Development Fund.
\end{acknowledgments}

\appendix

\section{On the \textgreek{Y}DO $I_{\ell}\left(\kappa\partial_{v}\right)$\label{sec:on-the}}

\begin{figure*}
\begin{centering}
\includegraphics[width=0.92\textwidth]{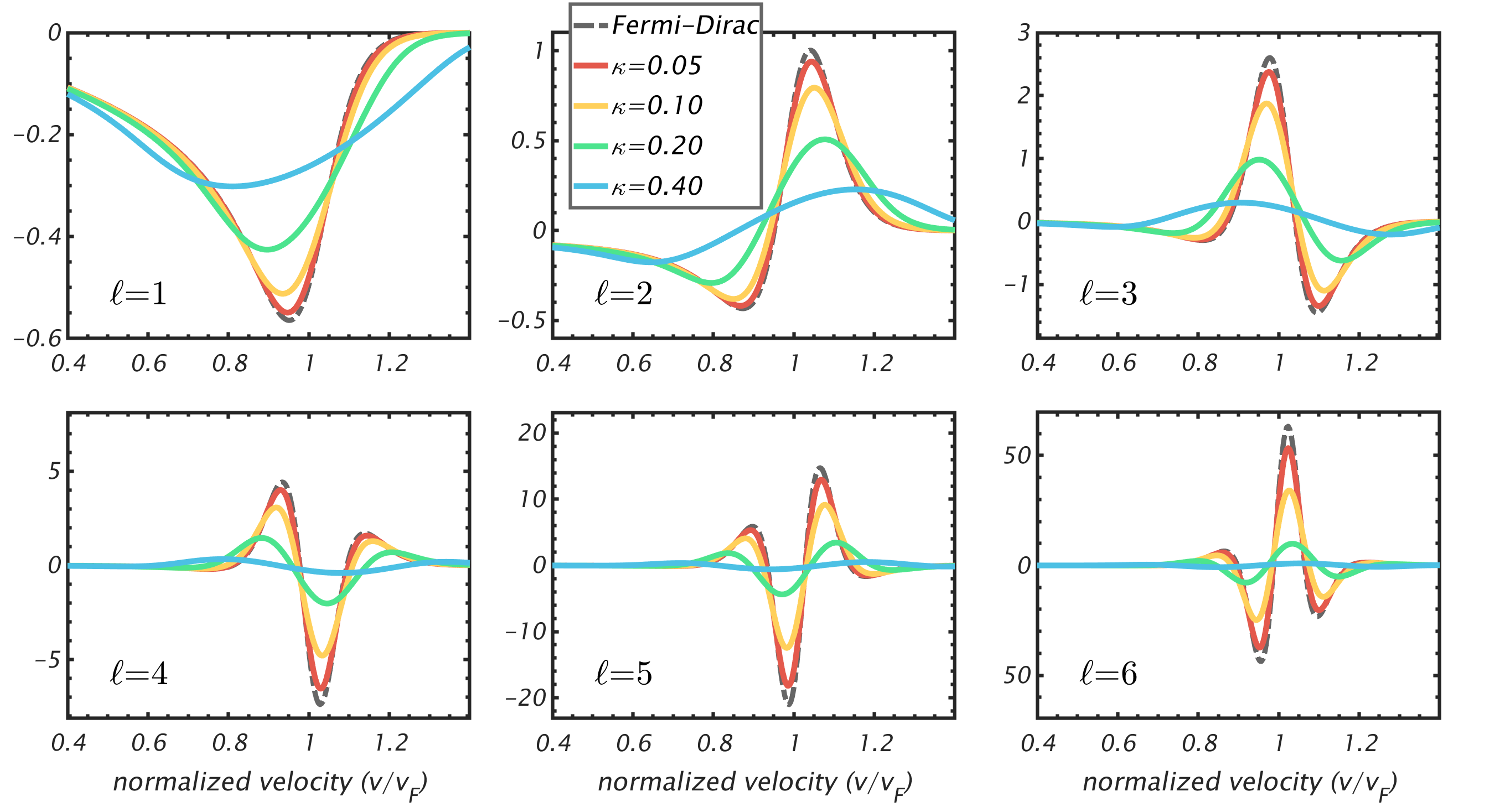}
\par\end{centering}
\caption{Numerical results of $I_{\ell}\left(\kappa\partial_{v}\right)F_{\perp}\left(v\right)/\kappa^{\ell}$
with $\ell=1\sim6$. The black dashed-lines are the right-hand-side
of \prettyref{eq:classical_Jf}.}
\label{fig:Jf_verify}
\end{figure*}

To prove \prettyref{eq:Jlf0}, we define $\hat{f}\left(\lambda\right)=\mathcal{F}^{-1}\left[f\left(v\right)\right]$
as the inverse Fourier coefficient of $f\left(v\right)$. Then we
have:

\begin{equation}
\begin{aligned}I_{\ell}\left(\kappa\partial_{v}\right)f\left(v\right)= & \ii^{-\ell}\int\d\lambda J_{\ell}\left(\kappa\lambda\right)\hat{f}\left(\lambda\right)\ee^{-\ii v\lambda}\\
= & \frac{\ii^{-\ell}}{2\pi}\iint\d\lambda\d v'J_{\ell}\left(\kappa\lambda\right)f\left(v'\right)\ee^{-\ii\left(v-v'\right)\lambda}\\
= & \ii^{-\ell}\mathcal{F}\left[J_{\ell}\left(\kappa\lambda\right)\right]\left(v\right)*f\left(v\right),
\end{aligned}
\label{eq:Iellf}
\end{equation}
where $*$ stands for convolution. And
\begin{equation}
\begin{aligned}\mathcal{F}\left[J_{\ell}\left(\kappa\lambda\right)\right]\left(v\right) & =\frac{1}{2\pi}\int_{-\pi}^{\pi}\mathcal{F}\left[\ee^{\ii\kappa\lambda\sin\theta}\right]\ee^{-\ii\ell\theta}\d\theta\\
 & =\mathrm{Rect}\left(\frac{v}{2\kappa}\right)\int_{-\pi}^{\pi}\delta\left(v+\kappa\sin\theta\right)\ee^{-\ii\ell\theta}\d\theta,
\end{aligned}
\label{eq:FJ}
\end{equation}
where
\[
\mathrm{Rect}\left(x\right)=\begin{cases}
1, & \text{if }\left|x\right|<\frac{1}{2},\\
0, & \text{otherwise},
\end{cases}
\]
is the unit rectangular function. Using
\begin{equation}
\begin{aligned}\delta\left(v+\kappa\sin\theta\right)= & \frac{1}{\sqrt{\kappa^{2}-v^{2}}}\left[\delta\left(\theta+\arcsin\frac{v}{\kappa}\right)\right.\\
 & \left.+\delta\left(\theta-\arcsin\frac{v}{\kappa}+\pi\,\mathrm{sign}\frac{v}{\kappa}\right)\right],
\end{aligned}
\end{equation}
then
\begin{equation}
\begin{aligned}\mathcal{F}\left[J_{\ell}\left(\kappa\lambda\right)\right]= & \frac{\mathrm{Rect}\left(v/2\kappa\right)}{\sqrt{\kappa^{2}-v^{2}}}\left(\ee^{\ii\ell\arcsin\frac{v}{\kappa}}+\ee^{\ii\ell\pi}\ee^{-\ii\ell\arcsin\frac{v}{\kappa}}\right)\\
= & \frac{2\mathrm{Rect}\left(v/2\kappa\right)}{\sqrt{\kappa^{2}-v^{2}}}\ee^{\ii\frac{\ell\pi}{2}}\mathbb{R}\mathrm{e}\left[\ee^{-\ii\frac{\ell\pi}{2}}\ee^{\ii\ell\arcsin\frac{v}{\kappa}}\right]\\
= & \frac{2\ii^{\ell}\mathrm{Rect}\left(v/2\kappa\right)}{\sqrt{\kappa^{2}-v^{2}}}\mathbb{R}\mathrm{e}\left(\frac{\kappa}{\sqrt{v^{2}-\kappa^{2}}-v}\right)^{\ell},
\end{aligned}
\label{eq:FJell}
\end{equation}
where we have used
\begin{equation}
\begin{aligned}\ee^{\ii\arcsin\frac{v}{\kappa}} & =\ii\frac{\sqrt{v^{2}-\kappa^{2}}+v}{\kappa}=\frac{\ii\kappa}{\sqrt{v^{2}-\kappa^{2}}-v}\end{aligned}
.
\end{equation}
Combining \prettyref{eq:Iellf} and \prettyref{eq:FJell}, we obtain
\prettyref{eq:Jlf0} in the main text:
\begin{equation}
\begin{aligned}I_{\ell}\left(\kappa\partial_{v}\right)f\left(v\right) & =\mathbb{R}\mathrm{e}\int_{-1}^{1}\frac{f\left(v-\kappa s\right)\d s}{\pi\sqrt{1-s^{2}}\left(\sqrt{s^{2}-1}-s\right)^{\ell}}\\
 & =\frac{\left(-\right)^{\ell}}{\pi}\int_{0}^{\pi}f\left(v-\kappa\cos\varphi\right)\cos\left(\ell\varphi\right)\d\varphi.
\end{aligned}
\label{eq:Jlf0-1}
\end{equation}
Alternatively, one can apply the sifting property of the $\delta$-function
in \prettyref{eq:FJ} to obtain the same result.

In small $\kappa$ limit, namely, the classical limit, we have
\begin{equation}
\lim_{\kappa\rightarrow0}\kappa^{-\ell}I_{\ell}\left(\kappa\partial_{v}\right)f\left(v\right)=\frac{1}{2^{\ell}\ell!}\partial_{v}^{\ell}f\left(v\right),\label{eq:classical_Jf}
\end{equation}
and we can make use of this feature to verify the numerical results.
Take the perpendicular Fermi-Dirac distribution \prettyref{eq:F_perp}
as an example, here are its first few derivatives:
\begin{equation}
\begin{aligned}\partial_{v}F_{\perp} & =CvP_{-\frac{1}{2}},\\
\partial_{v}^{2}F_{\perp} & =-C\left(v^{2}P_{-\frac{3}{2}}-P_{-\frac{1}{2}}\right),\\
\partial_{v}^{3}F_{\perp} & =-C\left(-v^{3}P_{-\frac{5}{2}}+3vP_{-\frac{3}{2}}\right),\\
\partial_{v}^{4}F_{\perp} & =-C\left(v^{4}P_{-\frac{7}{2}}-6v^{2}P_{-\frac{5}{2}}+3P_{-\frac{3}{2}}\right),\\
\partial_{v}^{5}F_{\perp} & =-C\left(-v^{5}P_{-\frac{9}{2}}+10v^{3}P_{-\frac{7}{2}}-15vP_{-\frac{5}{2}}\right),\\
\vdots
\end{aligned}
\end{equation}
where 
\begin{equation}
C=\frac{3}{4\pi v_{\mathrm{F}}^{3}},\quad P_{n}=\mathrm{Li}_{n}\left[-\ee^{\beta\left(\mu-\frac{m}{2}v^{2}\right)}\right].
\end{equation}
Numerical results of $\kappa^{-\ell}I_{\ell}\left(\kappa\partial_{v}\right)F_{\perp}\left(v\right)$
are presented in \prettyref{fig:Jf_verify}, with $\ell=1\sim6$,
and the dashed-lines stand for the classical operator $\partial_{v}^{\ell}F_{\perp}\left(v\right)/\left(2^{\ell}\ell!\right)$.
One can see that the $\Psi$DO \textbf{$I_{\ell}\left(\kappa\partial_{v}\right)$
}reduces to a differential operator as $\kappa$ tends to 0, which
means the non-local (quantum) \textgreek{Y}DO reduces to a local (classical)
derivation. 

\section{Derivation of the magnetized equilibrium Wigner function\label{sec:Derivation-of-the}}

We start from the definition of the Wigner function \citep{kremp2005quantum,kadanoff2018quantum}
\begin{equation}
f\left(\bx,\bv,t\right)=\int\frac{\d\omega}{2\pi}g^{<}\left(\omega,\frac{m\bv}{\hbar};t,\bx\right)
\end{equation}
where 
\[
g^{<}\left(\omega,\bq;t,\bx\right)=\int\d\tau\int\d\bxi\ee^{\ii\left(\bxi\cdot\bq-\omega\tau\right)}g^{<}\left(\tau,\bxi;t,\bx\right),
\]
and $g^{<}\left(\tau,\bxi;t,\bx\right)$ is the less Green function
$g^{<}\left(\bx_{2},\bx_{1};t_{2},t_{1}\right)$ in the Wigner coordinate:
\begin{equation}
\bxi=\bx_{2}-\bx_{1},\tau=t_{2}-t_{1},\bx=\frac{\bx_{2}+\bx_{1}}{2},t=\frac{t_{2}+t_{1}}{2}.
\end{equation}
In an uniform equilibrium system, $g^{<}$ is independent on $\bx$
and $t$. 
\begin{equation}
g^{<}\left(\omega,\bq\right)=\bar{f}^{<}\left(\omega\right)a\left(\omega,\bq\right),
\end{equation}
where 
\begin{equation}
\bar{f}^{<}\left(\omega\right)=\frac{1}{1+\ee^{\beta\left(\hbar\omega-\mu\right)}},
\end{equation}
and $a\left(\omega,\bq\right)$ is the spectrum function. In the RPA
limit, it is simply the Fourier component of the non-interactive single
particle Green function \citep{Ueta1992,MORGENSTERNHORING1976216},
which is
\begin{equation}
a\left(\omega,\bq\right)=2\ii\sum_{n=0}^{\infty}\frac{\left(-\right)^{n}\ee^{-q_{\perp}^{2}\ell_{B}^{2}}L_{n}\left(2q_{\perp}^{2}\ell_{B}^{2}\right)}{\omega-\hbar q_{\parallel}^{2}/2m-\left(n+\frac{1}{2}\right)\omega_{\mathrm{c}}}.
\end{equation}
Hence, by means of the residue theorem, we have
\begin{equation}
\begin{aligned}f\left(\frac{\hbar q_{\parallel}}{m},\frac{\hbar q_{\perp}}{m}\right)=2 & \ee^{-q_{\perp}^{2}\ell_{B}^{2}}\sum_{n=0}^{\infty}\left(-\right)^{n}L_{n}\left(2q_{\perp}^{2}\ell_{B}^{2}\right)\\
 & \times\bar{f}^{<}\left[\frac{\hbar q_{\parallel}^{2}}{2m}-\left(n+\frac{1}{2}\right)\omega_{\mathrm{c}}\right].
\end{aligned}
\end{equation}
With an additional normalizing factor $\mathcal{A}$, we obtain \prettyref{eq:Wigner-B}
in the main text.

\vspace*{1cm}

\bibliographystyle{unsrt}
\bibliography{ref}

\end{document}